\documentclass[aps,prl,twocolumn,groupedaddress,showpacs,showkeys,floatfix]{revtex4-1}
\usepackage[dvipdfm]{graphicx, color}
\usepackage{colordvi}
\usepackage{color}
\usepackage{float}
\usepackage{amsmath}
\usepackage{amssymb}
\usepackage{comment}
\usepackage{booktabs}
\renewcommand{\figurename}{Fig.}
\usepackage{lineno}

\begin{document}

\date{}

\title{Universal relationship in gene-expression changes for cells in steady-growth state}

\author{Kunihiko Kaneko}
\email[]{kaneko@complex.c.u-tokyo.ac.jp}
\affiliation{Research Center for Complex Systems Biology, Graduate School of Arts and Sciences, University of Tokyo,
Komaba, Tokyo, 153-8902, Japan}
\author{Chikara Furusawa}
\email[]{chikara,furusawa@riken.jp}
\affiliation{
Quantitative Biology Center (QBiC), RIKEN, 6-2-3 Furuedai, Suita, Osaka 565-0874, Japan}
\author{Tetsuya Yomo}
\email[]{yomo@ist.osaka-u.ac.jp}
\affiliation{
Department of Bioinformatics Engineering,
Graduate School of Information Science and Technology, 
Graduate School of Frontier Biosciences, Osaka University  and
Exploratory Research for Advanced Technology, Japan Science and Technology Agency, Suita, Osaka, Japan
}


\begin{abstract}
Cells adapt to different conditions by altering a vast number of components, which is measurable using transcriptome analysis. Given that a cell undergoing steady growth is constrained to sustain each of its internal components, the abundance of all the components in the cell has to be roughly doubled during each cell division event. From this steady-growth constraint, expression of all genes is shown to change along a one-parameter curve in the state space in response to the environmental stress. This leads to a global relationship that governs the cellular state: By considering a relatively moderate change around a steady state, logarithmic changes in expression are shown to be proportional across all genes, upon alteration of stress strength, with the proportionality coefficient given by the change in the growth rate of the cell. This theory is confirmed by transcriptome analysis of {\sl Escherichia Coli} in response to several stresses. 
\end{abstract}


\maketitle

{\bf Popular Summary}
{\sl 
Cells consist of a vast number of components whose concentrations are now measurable by means of transcriptome analyisis for gene expressions. 
Then, is it possible to extract biologically relevant features such as cellular growth, adaptation, and differentiation from such high-dimensional data? 
Can we uncover a universal law that governs across these high-dimensional data of gene expression levels? 
Here, recall that thermodynamics achieved a description by just few macroscopic variables from the motion of an immense number of molecules, 
by restricting our concern to thermal equilibrium. Of course, cells are not in equilibrium. Instead, they grow and divide, while keeping their 
concentrations of components at an approximately same level, in a steady-growth state. 
If we restrict our concern to such cells under a steady-growth condition, it implies that all the intracellular components are approximately doubled 
before cell division. From this constraint, a general law governing changes in gene expression during adaptation to environmental changes is 
derived theoretically; According to this law, changes in the expression of each gene are shown to be highly correlated, with a proportion 
coefficient determined by the growth rate of the number of cells; this is confirmed from transcriptome data of bacteria, 
{\it Escherichia coli} under different levels and types of environmental stresses. These correlated changes represent cellular homeostasis in 
response to environmental changes, set a constraint on high-dimensional changes in expression, represented by a single quantity, 
i.e., the cell growth rate, and facilitate a macroscopic description of cells during adaptation and evolution.
}

\section{Introduction}

Cell's internal state is now measurable with expression data on a few thousand genes using transcriptome analysis.
High-dimensional data on the gene expressions are gathered, depending on cells and environmental conditions.
In spite of the increase in the available data, however, it is sometimes difficult to extract biologically relevant characteristics from them, due to the complexity in gene expression network and dynamics.
Indeed, the common trend in the transcriptome analysis is to uncover a set of genes that specifically respond to specific environmental changes, while discarding other high-dimensional data that are gathered. Search for a simple law that governs a global change in expressions across genes has not seriously been attempted,

On the other hand, biologists are traditionally interested in macroscopic quantity such as activity, plasticity, and robustness\cite{Waddington,book}, 
even though these have involved qualitative, rather than quantitative, characteristics so far.
At this stage, then, it will be crucial to extract such macroscopic quantities from a vast amount of the expression data available using transcriptome analysis.
Here, the simplest candidate for such macroscopic quantity will be the growth rate in cell population.  Then, can we extract some universal relationship on global gene expression changes and connect it with a macroscopic (population) growth rate of cell?
 
In searching for such universal relationship, it will be relevant to restrict cell states of our concern, just as thermodynamics, the celebrated macroscopic phenomenological theory is established by restricting our concern to thermal equilibrium.  
Of course, a cell is not in a state of static equilibrium, but involves complex dynamics, and grows (and divides) in time. 
Thus we cannot apply the formulation in thermodynamics directly. 
However, we can instead follow the spirit in thermodynamics; we restrict our concern to a system with steady growth state and intend to extract a common law that should hold globally to such state. Considering that the cell keeps its internal state across cell divisions, it is expected that all the components grow with a common rate. 
As a consequence of such restriction, then, we may hope to uncover a universal relationship across changes in gene expressions. 
Indeed, in transcriptome analysis data, (e.g.,\cite{MA0,MA1,MA2}), existence of the correlation in the expression changes across a vast number of genes is suggested 
\cite{Marx,Alon,Braun,Matsumoto,Ying}, which are brought about through adaptation and evolution \cite{Bahler,Barkai,Lehner,Bahler2}. 

Here, we first analyze the transcriptome data in bacteria undergoing stress, to confirm a general relationship between global changes across expression of all genes. To explain such a general relationship in a cellular state, we study a general consequence imposed by a constraint of the cellular states achieved by restricting our concern only to cells that maintain steady growth, i.e., those cells that can grow and divide, retaining their state. 
Within this constraint, we derive a theoretical relationship of the changes in all components (i.e., expression levels of all genes) in response to stress. Following this theoretical framework, we then re-analyze transcriptome data to demonstrate the validity of our theoretical argument.

\section{Result}

{\bf Changes in gene expression under environmental stress conditions: experimental observations}

\begin{figure}[h!]
\begin{center}
\includegraphics[width=8cm]{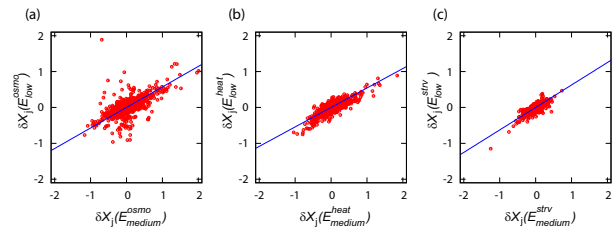}
\end{center}
\caption{Examples of the relationship between changes in gene expression $\delta X_j(E^a_{s_1})$ and $\delta X_j(E^a_{s_2})$ for genes in {\sl Escherichia Coli}. $\delta X_j$ represents the difference in the logarithmic expression level of a gene $j$ between the non-stressed and stressed conditions, where $s_1$ and $s_2$ represent two different stress strengths, i.e., low and medium. (a), (b), and (c) show the plot for $a=$ osmotic pressure, heat, and starvation stress, respectively. The fitted line is obtained by the major axis method, which is a least-square fit method that treats horizontal and vertical axes equally, and is usually used to fit bivariate scatter data\cite{Watson}. The slopes are 0.57, 0.54, and 0.62 for (a), (b), and (c), respectively. The expression data are obtained from \cite{Matsumoto}. Throughout the paper, we used the expression data of genes of which the expression levels under the three stress conditions as well as the original condition exceed a threshold ($X_i > -1.5$), in order to exclude inaccurate data (about 10\% of the total genes were discarded from the analysis).
}
\end{figure}

In \cite{Matsumoto}, transcriptome analysis of {\sl Escherichia Coli} under three environmental stress conditions, namely, osmotic stress, starvation, and heat stress, was carried out using microarrays. For each of these three conditions, three levels of stresses ($s=$ high, medium, low) were used, so that the absolute expression levels, represented by $x_j$ for $j$-th gene, are measured over a total of $3\times3$ conditions in addition to the original (stress-free) condition. To study behavior of cells under steady-growth conditions, cells were cultured for a sufficient period beyond the transient response to these stresses, after which gene expression levels were measured. Note that, throughout the paper, the point of interest is cellular behavior after recovery of the steady-growth state (which could be termed {\sl adaptation}, even though this does not necessarily imply the optimization of the growth rate or the genetic change). From these measurements, we calculated the change in gene expressions levels between the original state and that of a system experiencing environmental stress.

We investigated the difference in gene expression using a log-scale ($X_j=\log x_j$), that is $\delta X_j(E)=X_j(E)-X_j^O$ (i.e., $\log(x_j(E)/x_j^O)$) for genes $j$, where $E$ represents a given environmental condition, and $X_j^O$ represents the log-transformed gene expression level under the original condition. We adopted a logarithmic scale as changes in gene expression typically occur on this scale, and also as it facilitates comparison with the theory described below.

To characterize global changes in expression induced by these environmental stresses, we plotted the relationship between the differences in expression $(\delta X_j(E^a_{s_1}),\delta X_j(E^a_{s_2}))$ in Fig. 1a-c for $s_1=$ low and $s_2=$ medium, where $a$ is either osmotic, heat, or starvation stress. The relationship between all possible combinations of stresses and stress strengths are presented in Supplemental Fig. S1. For the same type of stress, $(\delta X_j(E^a_{s_1}),\delta X_j(E^a_{s_2}))$ correlates strongly over all genes, which suggests that the global trend in changes in expression levels can be represented by a small number of macroscopic variables.

\begin{figure}[h]
\begin{center}
\includegraphics[width=7cm]{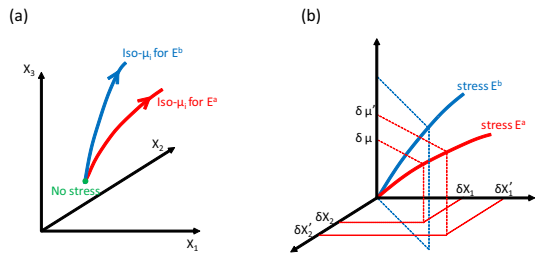}
\end{center}
\caption{
Schematic representation of our theoretical analysis: (a) Changes in gene expression in a high-dimensional state space, under a given environmental stress, follows a curve with $\mu_1=\mu_2=\cdots=\mu_M$, i.e., an iso-$\mu$ line. For different environmental conditions, the locus in the state space follows a different iso-$\mu$ line. (b) Changes in expression for each gene $1,2,\cdots$ is governed by the change in the growth rate $\delta \mu$. For different stress types, the change is shifted while governed by $\delta \mu$.
}
\end{figure}
	
\noindent
{\bf Theory for the steady-growth state}

To discuss changes in cellular state in response to environmental changes, we introduce a simple theory assuming a steady-growth state in a cell. When a cell grows at this steady state and reproduces itself, all the components it contains, e.g., the proteins that are expressed, have to be approximately doubled\cite{book,Zipf}. The abundance of each component increases at an almost equal rate over the time-scale of cell division; if the growth rates of some components were lower than that of others, the component would become diluted over time, and after some divisions, the component would be ``extinct", so that the cell state would not accommodate steady growth. For a cell to maintain the same internal state, all the components have to be synthesized at the same rate across cell divisions. This steady-growth condition has to be satisfied amidst the nonlinearity, complexity, and stochasticity of biochemical reactions.

Consider a cell consisting of $M$ chemical components, of which the synthesis allows it to grow and divide. In a cellular state under steady-growth conditions, the cell number increases exponentially over time, and thus each component within the cell also increases exponentially, as is expected from the autocatalytic nature of chemicals as a set of intracellular components. Hence, it is natural to assume that the abundance of components within the cell (as well as the cell volume) would generally grow exponentially over a cell division cycle. Then, the abundance of $i$-th component increases with $\exp(\mu_i t)$, over a cell division cycle, where $\mu_i$ is the growth rate of the component $i$. However, the steady-growth constraint under which the concentration of each component $i$ is maintained implies that $\mu_i=\mu$ for all components $i$. As $\mu_i$ is determined through the biochemical reactions in a cell, given from $M$-dimensional dynamics, the constraint $\mu_1=\mu_2=\cdots=\mu_M$ yields $M-1$ constraints on the $M$-dimensional state space (see Fig. 2a). After changes in the environment, there may be a transient period during which the cells have not yet attained this steady-growth state, but the steady state is likely to be attained over time, as long as the cell maintains all of its internal components. The growth rate $\mu$ itself is changed in response to a new condition, but the $M-1$ constraints $\mu_1=\mu_2=\cdots=\mu_M$ are preserved. Hence, over the long term, in response to environmental changes, the cell progresses along a one-dimensional curve in an $M$-dimensional state space of all components. This creates a general constraint on all gene expression levels.

Considering that $M$ represents a vast number (say $10^3-10^4$ protein species in typical cells), this reduction from $M$ to 1 is quite marked. Naturally, cells are not always in this steady-growth state. When a cell experiences different conditions, the growth rate of each component changes so that the concentration of each component is altered. Later, however, cells return to a steady growth-state with altered compositions of these components, somewhat analogous to the restrictions of thermal equilibrium state: when conditions within a system are changed, the temperature $T$ can become non-uniform. The temperature $T_i$ at a box $i$ can vary (sometimes on a microscopic scale, invalidating the existence of temperature itself), but after approaching equilibrium, all $T_i$'s are equal, so that a description using a few variables again becomes possible. Likewise, in our case, $\mu_i$ in the transient state could differ by component $i$, but after recovery of steady growth, all $\mu_i$'s are equal, allowing for a macroscopic description.

Next, we investigate the consequence of this constraint on steady growth. Consider the concentration $x_i(>0)$ of each component. Since each component $i$ is synthesized (or decomposed) in relationship to other components, the temporal change in the concentration of each component is represented as a function of the concentrations of the component itself and that of others, for instance by the rate-equation in chemical kinetics. Furthermore, each component, as well as the cell volume, grows at the rate $\mu$. Thus, the concentrations are diluted by this rate. Hence, the time-change of a concentration is given by

\begin{equation}
dx_i/dt=f_i(\{x_j\}) -\mu x_i.
\end{equation}

Now, the stationary state is given by a fixed point condition
\begin{equation}
x^*_i=f_i(\{x^*_j\}) /\mu
\end{equation}
for all $i$.

For the sake of convenience, let us denote $X_i=\log x_i$, and $f_i=x_i F_i$. Then, eq. (1) can be written as
\begin{equation}
dX_i/dt=F_i(\{X_j\}) -\mu,
\end{equation} 
with the corresponding fixed point solution
\begin{equation}
F_i(\{X^*_j \}) =\mu .
\end{equation}

In response to environmental changes, the growth rate $\mu$ itself changes, as does each concentration $x_i^*$; however, the $M-1$ condition requiring that $F_i(\{X^*_j \})$ is independent of $i$ for all $i=1,..,M$ has to be satisfied. Thus, a cell has to stay at a 1-dimensional curve in the $M$-dimensional space, under a given change in the environmental conditions (e.g., against changes in stress strength; see Fig. 2a). With an environmental change, all concentrations, $\mu$, and $\{X^*_j\}$ generally change, while the condition that $F_i(\{X^*_j \})$ is independent of $i$ is maintained as long as the cells continue steady-state growth.

We assume that all the components $i=1,2,..,M$ are retained after the change in environmental conditions, and that no new component (gene) emerges. Taken together, the cellular state is represented on an $M$-dimensional space. Now, consider intracellular changes in response to environmental changes as being represented by a set of continuous parameters $E^a$, which denote environmental changes under the stress condition $a$. For the moment, we omit the stress type $a$. With this parameterization $E$, the steady-growth condition leads to
\begin{math}
F_i(\{ X^*_j(E) \},E)=\mu(E).
\end{math}

We consider the parameter change from $E_0$ to $E$, where each $X_i^*$ changes from $X_i^*$ at $E_0$, to $X_i^*+\delta X_i$, which is accompanied by a change from $\mu$ to $\mu+\delta \mu$. Assuming a gradual change in the dynamics $x_i$, we introduce a partial derivative of $F_i(\{ X^*_j(E) \})$ by $X_j$ at $E=E_0$, which gives the Jacobi matrix $J_{ij}$. Now considering the condition under which the change is sufficiently small, and taking only the linear term in $\delta X_j$, we get
\begin{equation}
\sum_j J_{ij} \delta X_j (E) + \gamma_i \delta E =\delta \mu (E)
\end{equation}
with
\begin{math}
\gamma_i \equiv \frac{\partial F_i}{\partial E}.
\end{math}

Under the linear conditions we are concerned with, $\delta \mu \propto \delta E$, so that $\delta \mu= \alpha \delta E$ holds for a constant $\alpha$. Accordingly, we obtain
\begin{equation}
\delta X_j (E)=\delta \mu (E) \times \sum_i L_{ji}(1-\gamma_i/\alpha)
\end{equation}
\noindent  where $L=J^{-1}$.
Since the latter term on the right-hand side is independent of the magnitude of $E$, we simply have
\begin{equation}
\frac{\delta X_j(E)}{\delta X_j (E')}  =\frac{\delta \mu (E)}{\delta \mu (E')}
\end{equation}
\noindent over all $j$ (see Fig. 2b).

Hence, the change in the expression $X_j$ in response to external change is proportional over all components $j$ in this form. This provides a possible explanation for the observed transcriptome analysis shown Fig. 1. According to our theory, the proportion coefficient in the expression level should agree with the growth rate. Here, for each condition, the change in the growth rate $\delta \mu (E^a_s)$ was also measured. ($a$ is either osmotic, heat, or starvation stress).
In Fig. 3, we compared the slope of the changes in gene expression, i.e., the common ratio $\delta X_j(E^a_{s_1})/\delta X_j(E^a_{s_2})$, with $\delta \mu (E^a_{s_1})/\delta \mu(E^a_{s_2})$. The plot shows rather good agreement between these two. In this respect, the theory based on  steady-state growth and linearization of changes in stress  applies well to the transcriptome change.

\begin{figure}[h]
\begin{center}
\includegraphics[width=4.5cm,height=4.5cm]{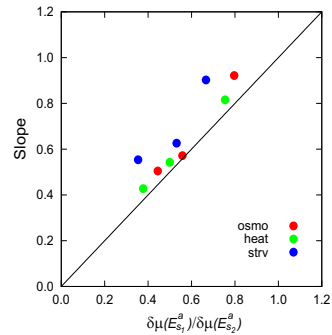}
\end{center}
\caption{
Relationship between the slope of the change in gene expression and the change in growth rate for the same stress types. The abscissa denotes $\delta \mu (E^a_{s_1})/\delta \mu (E^a_{s_2})$, while the ordinate is the slope in $(\delta X_j(E^a_{s_1}),\delta X_j(E^a_{s_2}))$. The red, green, and blue dots represent osmotic, heat, and starvation stress data, respectively, while the pair $s_1, s_2$ runs over different strengths of the same stress type.
}
\end{figure}

\noindent
{\bf Changes in gene expression across different types of stresses}

So far, we have compared the expression levels across different strengths for the same type of stress. However, expression changes can also be compared across different stress conditions. Interestingly, the genome-wide correlation of expression levels is not restricted to a change in the same stress condition. In Fig. 4, $(\delta X_j(E^a_{high}),\delta X_j(E^b_{high}))$, which plots expression changes across different stress conditions $a\neq b $ ($=$ either starvation, heat, or osmotic stress), correlation is still observed, even though there are more genes that deviate from the common proportionality, leading to lower correlation coefficients, as compared with the correlations observed under the same stress conditions.

The correlation is also discernible for other choices of $s_1,s_2$, as shown in Supplemental Fig. S1, where all the correlation diagrams of $(\delta X_j(E^a_{s_1}),\delta X_j(E^b_{s_2}))$ across all possible stress conditions are plotted.
Note that such proportionality across genes has also been suggested for several experiments, over different environmental conditions\cite{Alon,Braun}. 
The finding of correlation, even with reduced proportionality, implies a common trend in changes in expression across many genes, which is not necessarily the result of a given stress condition, but is a concept that holds across different environmental conditions.

Since gene expression dynamics are very high-dimensional, this correlation suggests the existence of a strong constraint to adaptive changes in expression dynamics. Below, we discuss the theoretical origin of this correlation. In eqs. (5)-(6), the environmental change $E$ is no longer represented by a scalar variable, but the environmental change involves a different direction, so that $\gamma_i^a$ and $\alpha^a$ depend on the type of environmental (stress) condition $a$. Hence, instead of eq. (7), we get
\begin{equation}
\frac{\delta X_j (E^a)}{\delta X_j (E^b)}  =\frac{\delta \mu (E^a)}{\delta \mu (E^b)}
\frac{ \sum_i L_{ji}(1-\gamma_i^a/\alpha^a) } {\sum_i L_{ji} (1-\gamma_i^b/\alpha^b)}
\end{equation}
Here, the right-hand side (RHS), in general, depends on each gene $j$. This could blur the proportionality in $(\delta X_j (E^a), \delta X_j (E^b) )$ over all genes.
In the following case, however, the dependence of the RHS on $j$ is relaxed, to support approximate proportionality as indicated in Fig. 4.
When $\gamma_i^a$ and $\gamma_i^b$ are independent of $i$, which we denote as $\gamma^a$ and $\gamma^b$, respectively, the RHS is reduced to
\begin{equation}
\frac{\delta \mu (E^a)}{ \delta \mu (E^b)} \frac{ (1-\gamma^a/\alpha^a)}{ (1-\gamma^b/\alpha^b)},
\end{equation}
so that the common proportionality of the change in expression holds, while the proportion coefficient is shifted from a simple ratio between the growth-rate changes $\delta \mu$.

Sometimes, environmental changes affect all processes, globally. For example, if temperature or nutrient resources are increased, the synthesis (or decomposition) rates of all reaction processes are amplified across the board.

Of course, there are some genes for which $\gamma^E_i$ deviates from the above common value. If the number of such genes with a specific response is small (and$/$or, its influence on other genes is small, i.e., the Jacobi matrix is sparse), then the contributions from genes with a common $\gamma^E$ value makes up the major portion of the summation in the RHS of eq. (9). If we neglect the minor contributions from a few specific genes, common proportionality could generally be maintained. Indeed, only a limited number of specific genes are expected to respond directly to environmental changes.

According to this approximation, the proportion coefficient $\delta X_j(E^a)/\delta X_j(E^b ) $ deviates from $\delta \mu (E^a)/\delta \mu(E^b)$ by the factor $(1-\gamma^a/\alpha^a)/(1-\gamma^b/\alpha^b)$. Note that this correction in the proportion coefficient depends only on the type, but not on the strength of each stress.

We examined this point from the transcriptome data analyzed here, by plotting the proportion coefficient in $\delta X_j(E^a)/ \delta X_j(E^b)$ versus $\delta \mu (E^a)/\delta  \mu(E^b)$ in Fig. 5. The correlation between $\delta X_j$ and the growth rate in this figure also exists across different stress conditions. Additionally, the coefficient $\delta X_j(E^a)/ \delta X_j(E^b)$ is roughly proportional to $\delta \mu (E^a)/ \delta \mu (E^b)$ with a proportion coefficient that is mainly determined by the pair of stress types, over different strengths.

Undeniably, the proportionality over different stress types is not optimal. 
Indeed, existence of gene-specific dependence $\gamma^a_i$ leads to scattering in $(\delta X_j(E^a), \delta X_j(E^b))$ (for $a\neq b$) 
around the common proportionality by genes, and  
there are more genes that deviate from the common proportionality for $a\neq b$ 
than those for $a=b$ ( compare Fig.4 with Fig.1), 
so that the estimation of the proportion coefficient in Fig.5 is not so reliable especially for those with lower correlation coefficient.

\begin{figure}[h]
\begin{center}
\includegraphics[width=8cm]{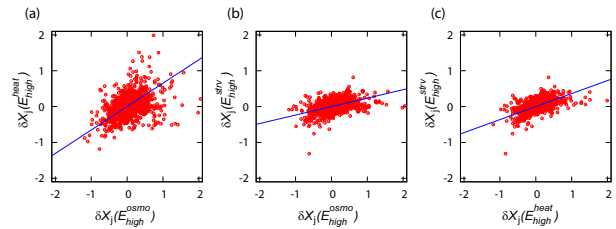}
\end{center}
\caption{
Examples of the relationship between changes in gene expression $\delta X_j(E^a_{s_1})$ and $\delta X_j(E^b_{s_2})$ for different stress types. The combination of stresses $(a, b)$ is (osmotic, heat) for (a), (osmotic, starvation) for (b), and (heat, starvation) for (c), respectively. The strengths of the stress $s_1$ and $s_2$ are fixed as high in these figures. The slopes are 0.65, 0.24, and 0.36 for (a), (b), and (c), respectively, while the correlation coefficient for each data is 0.40 (a), 0.43 (b) and 0.54(c). The relationships between the changes in gene expression for all possible combinations are presented in Supplemental Fig. S1. The fitted line is obtained by the major axis method as described in Fig. 1.
}
\end{figure}

\begin{figure}[h]
\begin{center}
\includegraphics[width=9cm]{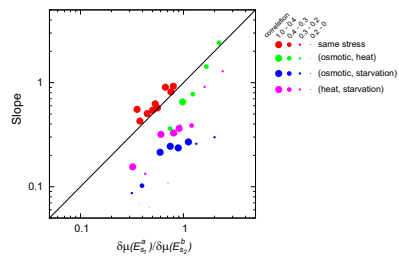}
\end{center}
\caption{
Relationship between the slope of change in gene expression and change in growth rate for different stress types.
The abscissa denotes $\delta \mu (E^a_{s_1})/\delta \mu (E^b_{s_2})$, while the ordinate is the slope in $(\delta X_j(E^a_{s_1}),\delta X_j(E^b_{s_2}))$. The red dots represent data for the same stress types, while the green, blue, and purple dots show combinations of different stress types $(a, b) =$(osmotic, heat), (osmotic, starvation), and (heat, starvation), respectively. The pair $s_1, s_2$ runs over different strengths of all stress combinations. The size of the dots represents the correlation coefficient between $\delta X_j(E^a_{s_1})$ and $\delta X_j(E^b_{s_2})$. A lower correlation indicates that the fit of the slope may be less accurate.
}
\end{figure}

\section{Discussion}

We have shown here that steady growth conditions lead to a global constraint over all gene expression patterns. With a few additional assumptions, the proportionality in the change in expression across genes can be derived, in which the proportion coefficient is mainly governed by the change in the growth rate. These theoretical predictions were compared with several bacterial gene expression experiments, with approximate agreement.

The correlation with the growth rate is also interpreted by neglecting the direct environment dependence in $F_i(\{X_j(E )\},E)$, i.e., by replacing it with $ F_i(\{X_j (E ) \})$ for most genes $j$. In other words, external environmental changes trigger changes in the levels of some components $\{ x_m(E) \}$, which introduces a change in the growth rate $\mu(E)$. For the stationary state, only the condition $F_i(\{ x_j(E) \})= \mu(E)$ is considered. 
With this approximation, the term for direct environmental changes $\gamma_i=\partial F_i/\partial E$ is neglected, and eq. (7) follows directly, so that growth rate changes determine gene expression changes globally. Indeed, the experimental data may suggest that the growth-rate makes the major contribution to changes in gene expression.

This dominancy of growth-rate is, however, imperfect, so that the environment-specific term $\gamma^a_i$  has to be taken into account to compare expression of genes across different stress conditions. In our simple approximation that neglects gene dependence of $\gamma^a_i$, $\gamma^a/\alpha^a$ represents the degree of the direct influence of the environment on gene expression dynamics, as compared with the influence on the growth rate.

As another possible estimate of this factor $\gamma^a/\alpha^a$, we directly measured the variance of changes in expression across genes, i.e., $<(\delta X_j(E) -<\delta X_j(E)>)^2>$, where $<\cdots>$ is the average over all genes (see Supplemental Fig. S2). According to eqs. (8) and (9), this factor grows in proportion to $(1-\gamma^a/\alpha^a)^2 \delta\mu^2$ (in addition to the variances determined by the Jacobi matrices, which are independent of environmental stress $a$). As shown in Supplemental Fig. S2, the factor $(1-\gamma^a/\alpha^a)$ decreases in the order of $a=$ osmotic stress, starvation, and heat stress. Indeed, the deviation from $\delta \mu^a/\delta \mu^b$ in Fig. 5 is consistent with the above order of $(1-\gamma^a/\alpha^a)$.

Furthermore, the environment-specific response of gene expression $\gamma^a_i$ generally depends on each gene $i$. Here, genes that show specific responses to a given environment $E^a$ may be few, while most others may not be influenced directly by the environment; their expression levels may be mostly determined by the homeostatic growth condition $F_i\{ X_j(E) \})= \mu(E)$. Distinguishing such homeostatic genes from those that show specific responses to individual environmental stresses will be important as the next step in the statistical analysis of adaptation.

In eq.(1), we have not assumed any specific form with particular dependence upon some protein species.  In recent study, specific dependence of the fraction of ribosomal proteins upon the growth rate is discussed by adopting a description by few degrees of protein groups \cite{Hwa}. It will be interesting to introduce some specific genes in our formulation while keeping high-dimensional expression dynamics.

It is also interesting to note that gene expression changes $\delta X_j$ across genes correlate between environmental and genetic perturbations. In fact, Ying et al.\cite{Ying} measured the changes in gene expression induced by the environmental perturbation $E^{env}$, and the genetic perturbation $E^g$ induced by external reduction of several genes. Again, they observed a strong correlation between $\delta X_j(E^{env})$ and $\delta X_j(E^g)$ across genes (see Fig. 5 of \cite{Ying}).

Indeed, our theory can also be applied to adaptive evolution, in which growth rate is first reduced by encountering a novel environment $E^{env}$, and then recovers by genetic changes $E^{g}$ through evolution, so that $\delta \mu (E^{g})/\delta \mu (E^{env})<1$. According to our relationship, changes in gene expression levels introduced by a new environment is reduced through adaptive evolution, i.e., $\delta X_i(E^{g})/\delta X_i(E^{env}) \sim  \delta \mu (E^{g})/\delta \mu (E^{env}) <1$, as discussed. In other words, there is a common homeostatic trend for the expression of most genes to return to the original level, as extensively observed experimentally\cite{Marx,heat}.

A few issues should be considered prior to the application of the theory presented here. First, it must be assumed that the components continue to exist, and that novel components do not appear. Under this condition, the postulate for common $\mu_i$ generally holds, even though the linear approximation does not. However, even if some components do become extinct or novel components emerge, the constraint may still exist for other components, and the proportionality relationship Eq. (7) holds approximately, as long as the influence of the extinct or emerging components is limited.

Second, it is assumed that the fixed-point of eq. (1) is not split by bifurcation. When bifurcation occurs, we can apply our theory along each branch (under the condition that the inverse Jacobi matrix exists), but direct comparisons cannot be made across different branches. Moreover, in some cases, the attractor of the expression level is not a fixed-point, but is an oscillatory state. However, as long as the oscillation period is shorter than the cell division time, one can use the average $\mu_i$ of the period, instead of $\mu_i$, leaving the present argument valid.

Third, we adopted a linearization approximation to obtain Eq. (7). For larger changes in external conditions, there will be a gene-specific correction to the linear relationship Eq. (6). However, linearization is adopted after taking the logarithm of gene expression levels, so that the size of $\delta X_i$ may not be so restrictive when seen in the original scale of gene expression $x_i$. Indeed, the agreement with the theory shown in Figs. 1 and 3 for the same stress indicates that the linearization approximation is valid, even though the growth rate is reduced to less than half of the original.

The present theory facilitates description of a cellular system with only few macroscopic variables, for characterization of adaptation and evolution. Furthermore, our theory with regards to common $\mu_i$ can be applied to any system of stationary growth. As presented, each element $i$ represents a replicating molecule within a cell, but, similarly, we can apply our theory by using such an element to describe cells of different types within an organism. Alternatively, macroscopically, one can assign an element as a population of each species in a stationary ecosystem. The multi-level constraint of the steady-growth condition across a hierarchy is an important concept for elucidating global relationships in complex-systems biology\cite{book}.

{\bf Acknowledgments.}
This work was supported in part by platform for Dynamic Approaches to the Living Systems from MEXT.

%


\vspace{7cm}

\makeatletter
\renewcommand{\thefigure}{S\@arabic\c@figure}
\makeatother

\vspace{4cm}
{\Large Supplemental Figures:  Kaneko et al.}

\begin{figure*}[htbp]
\begin{center}
\includegraphics[width=12cm]{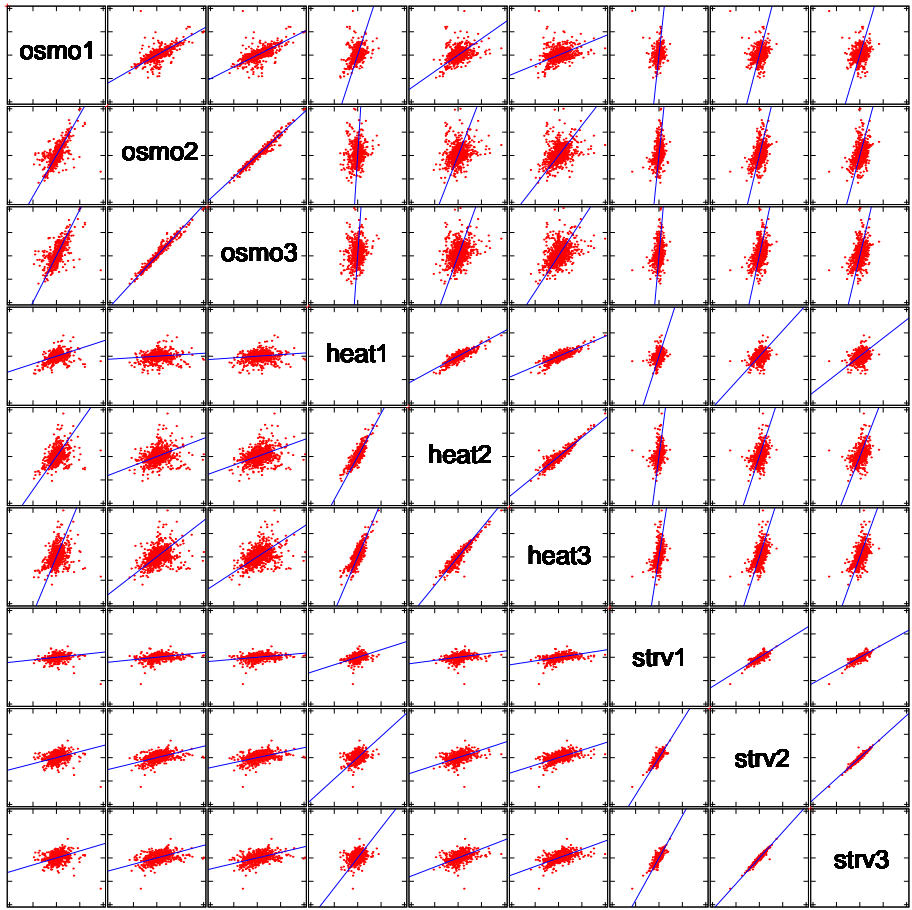}
\end{center}
\caption{ {\bf Supplementary Fig.1}:
Relationship between gene expression changes $\delta X_j(E^a_{s_1})$ and $\delta X_j(E^b_{s_2})$ for all possible combinations of stress types and stress 
strength, from the data in[16].  The blue fitted lines are obtained by the major axis method.
}
\end{figure*}

\begin{figure*}[htbp]
\begin{center}
\includegraphics[width=10cm]{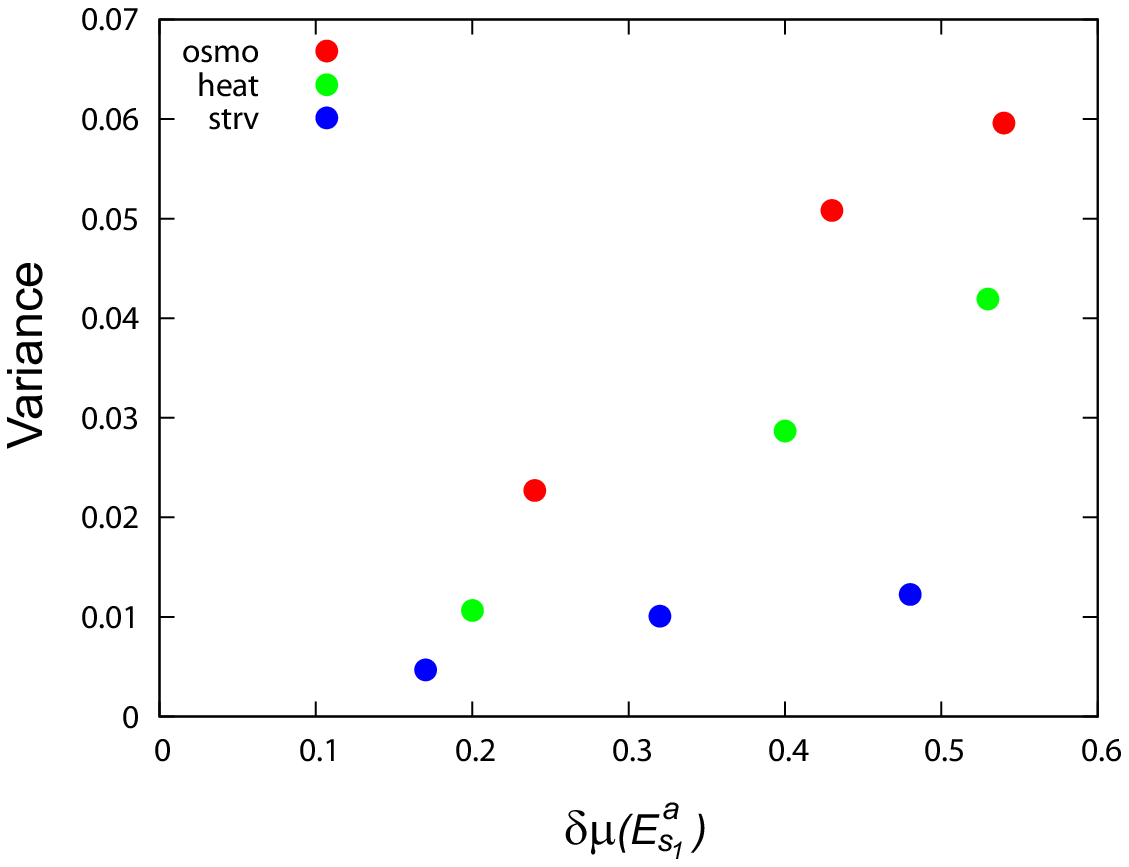}
\end{center}
\caption{{\bf Supplementary Fig.2}:
The variance of $\delta X_j(E^a_{s1})$ over all genes as a function of $\delta \mu(E^a_{s1})$, from the data in [16].
The variance 
of the expression change over genes 
is computed as
$<(\delta X_j(E^a_{s_1}) -<\delta X_j(E^a_{s1})>)^2>$, where $<\cdots>$ is the average over all genes.
The red, green, and blue dots denote $a=$ osmotic pressure, heat, and starvation stress, respectively.
}
\end{figure*}


\begin{thebibliography}{999}


\bibitem{MA0}
Eisen, M. B., et al. (1998) Cluster analysis and display of genome-wide expression patterns. {\sl Proc Nat Acad Sci USA} 95(25):14863-14868.

\bibitem{MA1}
Gunasekera TS, Csonka LN, Paliy O (2008) Genome-wide transcriptional responses of Escherichia coli K-12 to continuous osmotic and heat stresses. {\sl J Bacteriol} 190(10):3712-3720.

\bibitem{MA2}
Richmond CS, Glasner JD, Mau R, Jin H, Blattner FR (1999) Genome-wide expression profiling in Escherichia coli K-12. {\sl Nucleic Acids Res} 27(19):3821-3835

\bibitem{Zipf}
Furusawa C, Kaneko K (2003) Zipf's law in gene expression. {\sl Phys Rev Lett} \textbf{90,}:
088102 (2003).

\bibitem{Waddington}
Waddington CH (1957) {\sl The Strategy of the Genes} (Allen \& Unwin, London).

\bibitem{book}
Kaneko K (2006) {\sl Life: An Introduction to Complex Systems Biology} (Springer, Heidelberg and New York).

\bibitem{Bahler}
Chen D, et al. (2003) Global transcriptional responses of fission yeast to environmental stress. {\sl Mol Biol Cell} 14(1):214-229.

\bibitem{Barkai}
Bergmann S, Ihmels J, Barkai N. (2003) Similarities and differences in genome-wide expression data of six organisms. {\sl PLoS Biol} 2(1): e9.

\bibitem{Laundry}
Landry CR, Lemos B, Rifkin, SA, Dickinson WJ, Hartl DL (2007) Genetic properties influencing the evolvability of gene expression. {sl Science} 317: 118.

\bibitem{Lehner}
Lehner B, Kaneko K. (2011) Fluctuation and response in biology. {\sl Cell Mol Life Sci} 68(6):1005-1010.

\bibitem{Bahler2}
Marguerat S, et al. (2012) Quantitative analysis of fission yeast transcriptomes and proteomes in proliferating and quiescent cells. {\sl Cell} 151(3):671-683.

\bibitem{Braun}
Stern S, Dror T, Stolovicki E, Brenner N, Braun E (2007) Genome-wide transcriptional plasticity underlies cellular adaptation to novel challenge. {\sl Mol Syst Biol} 3(1):106.

\bibitem{Alon}
Keren L, et al. (2013) Promoters maintain their relative activity levels under different growth conditions. {\sl Mol Syst Biol} 9(1):701.

\bibitem{Hwa}
Scott M.,Gunderson C.W.,,Mateescu E.M., Zhang Z., Hwa T. (2010),
Interdependence of cell growth and gene expression: origins and consequences.
{\sl Science 330}, 1099-1102

\bibitem{Marx}
Carroll SM, Marx CJ (2013) Evolution after introduction of a novel metabolic pathway consistently leads to restoration of wild-type physiology. {\sl PLoS Genet} 9:e1003427

\bibitem{Matsumoto}
Matsumoto Y, Murakami Y, Tsuru S, Ying BW, Yomo T (2013) Growth rate-coordinated transcriptome reorganization in bacteria. {\sl BMC Genomics} 14(1): 808.

\bibitem{Ying}
Ying BW, Seno S, Kaneko F, Matsuda H, Yomo T (2013). Multilevel comparative analysis of the contributions of genome reduction and heat shock to the Escherichia coli transcriptome. {\sl BMC Genomics} 14(1): 25.

\bibitem{heat}
Kishimoto T, et al. (2010) Transition from positive to neutral in mutation fixation along with continuing rising fitness in thermal adaptive evolution. {\sl PLoS Genet} 6(10):e1001164

\bibitem{Watson}
Watson DI, Wright IJ, Falster DS, Westoby M (2006) Bivariate line-fitting methods for allometry. {\sl Biol Rev} 81(2): 259-291.


\end{thebibliography}
\end{document}